\documentclass[runningheads]{llncs}
\usepackage{graphicx}
\usepackage{multirow}
\usepackage{color}
\begin{document}
\title{Are fast labeling methods reliable? A case study of computer-aided expert annotations on microscopy slides}
\titlerunning{Are fast labeling methods reliable?}
%


\author{Christian Marzahl\inst{1,2} \and
Christof~A.~Bertram \inst{3} \and 
Marc Aubreville\inst{1} \and 
Anne Petrick \inst{3}\and 
Kristina Weiler\inst{4}\and 
Agnes C. Gl{\"a}sel \inst{4}\and 
Marco Fragoso \inst{3}\and 
Sophie Merz \inst{3}\and 
Florian Bartenschlager \inst{3}\and 
Judith Hoppe \inst{3}\and 
Alina Langenhagen \inst{3}\and 
Anne Jasensky \inst{3}\and 
J{\"o}rn Voigt\inst{2} \and 
Robert~Klopfleisch \inst{3} \and 
Andreas Maier\inst{1}}
\authorrunning{C. Marzahl et al.}
%
\institute{Pattern Recognition Lab, Department of Computer Science, Friedrich-Alexander-Universit{\"a}t \and
Research \& Development, EUROIMMUN Medizinische Labordiagnostika AG \and
Institute of Veterinary Pathology, Freie Universit{\"a}t Berlin, Germany  \and
Department of Veterinary Clinical Sciences, Clinical Pathology and 
Clinical Pathophysiology, Justus-Liebig-University Giessen, Germany
}
%
\maketitle              
\begin{abstract}
Deep-learning-based pipelines have shown the potential to revolutionalize microscopy image diagnostics by providing visual augmentations and evaluations to a trained pathology expert. However, to match human performance, the methods rely on the availability of vast amounts of high-quality labeled data, which poses a significant challenge. To circumvent this, augmented labeling methods, also known as expert-algorithm-collaboration, have recently become popular. However, potential biases introduced by this operation mode and their effects for training deep neuronal networks are not entirely understood. 
This work aims to shed light on some of the effects by providing a case study for three pathologically relevant diagnostic settings. 
Ten trained pathology experts performed a labeling tasks first without and later with computer-generated augmentation. To investigate different biasing effects, we intentionally introduced errors to the augmentation. Furthermore, we developed a novel loss function which incorporates the experts' annotation consensus in the training of a deep learning classifier.
In total, the pathology experts annotated 26,015 cells on 1,200 images in this novel annotation study. Backed by this extensive data set, we found that the consensus of multiple experts  and the deep learning classifier accuracy, was significantly increased in the computer-aided setting, versus the unaided annotation. However, a significant percentage of the deliberately introduced false labels was not identified by the experts. Additionally, we showed that our loss function profited from multiple experts and outperformed conventional loss functions. At the same time, systematic errors did not lead to a deterioration of the trained classifier accuracy. Furthermore, a deep learning classifier trained with annotations from a single expert with computer-aided support can outperform the combined annotations from up to nine experts. 

\end{abstract}
\section{Introduction}

The field of computer vision strongly relies on the availability of high quality, expert-labelled image data sets to develop, train, test and validate algorithms. The availability of such data sets is frequently the limiting factor for research and industrial projects alike. This is especially true for the medical field, where expert resources are restricted due to the high need for trained experts in clinical diagnostics. Consequently, the generation of high-quality and high-quantity data sets is limited and there is a growing need for a highly efficient labeling processes. To explore the potential of reducing the expert annotation effort while maintaining expert-level quality, we reviewed a method called expert-algorithm collaboration in three types of data sets. In this approach, experts manually correct labels pre-computed typically by a machine learning-based algorithm. While there are numerous successful applications of crowd-sourcing in the medical field~\cite{0000-00} crowd-algorithm collaboration or expert-algorithm collaboration has been applied rarely and only in order to solve highly specific tasks. Some examples are: Maier-Hein et al.~\cite{0000-01} on endoscopic images, Ganz et al.~\cite{0000-02} on MR-based cortical surface delineations or Marzahl et al.~\cite{0000-10} on a pulmonary haemorrhage cytology data set.
There is no previous research regarding the best way to apply expert-algorithm collaboration, or its challenges and limitations. Therefore, this study aims to investigate the following questions: First, is the expert-algorithm collaboration performance independent from the type of task? Second, can the findings in~\cite{0000-10} regarding the performance gains for some specific tasks also be applied to different types of medical data sets? Third, is there a bias towards accepting pre-computed annotations? Finally, is a loss function that tries to incorporate the varying experts' annotation quality better suited to train a deep learning classifier than conventional state-of-the-art loss functions?
To test our hypothesis, ten medical experts participated in our expert-algorithm collaboration study on a mitotic figure detection, asthma diagnosis, and pulmonary haemorrhage cytology data set. 

\section{Material and methods}

For our experiments, we selected three different types of medical object detection data sets.
First, as an example of a non-challenging classification task, we chose the microscopic diagnosis of equine asthma. Equine asthma is diagnosed by counting five types of cells (eosinophils, mast cell, neutrophils, macrophages, lymphocytes) and calculating their probability of occurrence. The cells are visually clearly distinguishable from each other due to their morphology, size and structure. The data set~\cite{doi:10.1111/vcp.12623} was created by digitisation of six May-Grunwald Giemsa stained cytocentrifugated equine bronchoalveolar lavage fluids. 
Second, equine exercise-induced pulmonary haemorrhage (EIPH) is diagnosed by counting hemosiderin-laden macrophages, which can be divided into five groups according to Golde et al.~\cite{0000-07}. In contrast to equine asthma, the cells are stained with Prussian Blue or Turnbull's Blue in order to visualise the iron pigments contained in the hemosiderin. The grading task, however, is particularly challenging because the hemosiderin absorption is a continuous process which is mapped to a discrete grading system. 
The last task we considered was the detection of rare events on microscopy images. As such, the identification of mitotic figures (i.e., cells undergoing cell division) is a prominent example used in the vast majority of tumor grading schemes and known to have high inter-rater disagreement \cite{Meyer:2005cl,doi:10.1177/0300985819890686}. Due to the rareness of mitotic figures in histopathological specimens, this represents a challenging task with high demands on concentration and expert knowledge, and is thus a very suitable candidate for visual augmentation in a decision support system in a clinical diagnostic setting. Whole slide images (WSI) containing 21 tumor cases with mixed grade were selected from the training set of a publicly available data set of canine cutaneous mast cell tumor \cite{Bertram_SciData}, which represents hematoxylin and eosin stained specimen of various tumor grades at a resolution of $0.25 \frac{\mu m}{\textrm{px}}$. 
Experts were requested to annotate mitotic figures they identified as required by the commonly used grading scheme \cite{Kiupel:2011du}. 
To quantify the quality of the experts' annotations, we calculated the mean intersection over union (mIoU) and trained a deep learning classifier using a custom, novel loss function.

\subsection{Patch selection}

For labelling and training the data set was subdivided into patches. The patch selection process aims to represent the variability of the data set as carefully as possible while providing a realistic workflow of integrating pre-computed annotations into clinical routine. Therefore, we selected twenty patches for EIPH, Asthma and mitotic figures:
For EIPH, we used twenty algorithmically chosen patches which had to fulfill the following criteria: each patch had to cover as many grades as possible, the two staining types had to be represented equally, at least one patch was sampled from each WSI and as recommended for grading by Golde et al. ~\cite{0000-07} around 300 hemosiderophages were visible on all patches combined.
The twenty asthma patches were selected on the condition that all cell types are represented equally. Additionally, the patches had to have the highest visual quality possible.
From the WSI of the mitotic figure data set, a board-certified pathologist selected a region of interest spanning 10 high power fields (10 HPF, total area=$2.37\,mm^2$), as done in a diagnostic setting. This restriction of area was performed in order to reduce known selection bias for this study \cite{aubreville2019field}. 

\subsection{Label generation}
Besides labels considered as ground truth from the respective data sets, we deliberately included modifications to these labels to investigate their effects and potential biases in the resulting data set generated by the experts annotating in an expert-algorithm collaborative setting.
Matching the tasks, we introduced different annotation errors:
\paragraph{\textbf{Equine asthma data set}}
For the equine asthma data set, we randomly changed the cell type of 15 cells on five of the images. Additionally, on a separate set of five images we removed the annotation of one cell, thus introducing a missing annotation.
\paragraph{\textbf{EIPH scoring}} For the EIPH grading task, we also introduced missing cells on five of the images. On a distinct set of five further images, we increased the grade of each of the cells by one. Finally, five images contained in total ten standard artifacts that could be generated by a typical object detection pipeline, such as false annotations (where no relevant cell was present) or multiple annotations.

\paragraph{\textbf{Mitotic figures}} For the mitotic figure identification task, we removed 20\,\% of all mitotic figures (resulting in 149 missing objects) present in the data set and added the same amount of false mitotic figures. To further understand the effects introduced by this, the mitotic figures were categorized by a CNN-based classifier w.r.t. their model score of being a mitotic figure. We did this in order to provide easy to spot false annotations, hard to distinguish candidates, easily recognizable mitotic figures (as per model score), and hard to identify objects. 
We grouped the mitotic figures accordingly as 0: easy (n=49), 1: medium (n=50) and 2: hard (n=50) for the fake mitotic figure task. The cutoff thresholds were chosen at $p_0\leq0.2$, $0.2< p_1 \leq0.4$ and $p_2>0.4$, respectively, where $p$ is the final model score in a binary classification experiment.
For the deleted true mitotic figures, we also performed a selection according to their group, where group 0 represented the hard to spot mitotic figures (n=49), 1 the medium level (n=59) and 2 the easy to spot mitotic figures (n=50), as given by the model's score. To define the groups, we randomly selected according to the thresholds $p_0\leq0.33$, $0.33 < p_1 \leq0.66$ and $p_2>0.6$.

While participants were informed that the proposed labels would be computer-generated, they were unaware about the distinctively introduced false labels. This allowed us to investigate the effects of error propagation introduced by our systematic labeling flaws.

\subsection{Labelling platform}

We used the open-source platform EXACT~\footnote{\url{https://github.com/ChristianMarzahl/Exact}} to host our experiments. 
Anonymity was ensured by using a secure server and by removing any personal information from the files' meta-data.

\subsection{Label experiment design}

We designed our experiments with the intent to assess the effect of computer-aided annotation methods on medical experts. For that purpose, we created three modes on the EXACT server for each of the three types of data sets (EIPH, Mitotic figures, Asthma). The first mode is the training mode which can be used to get familiar with the user interface and the EXACT features. The training was supported by providing YouTube~\footnote{\url{https://youtu.be/XG05RqDM9c4}} videos describing the user interface as well as the tasks. The second mode is the annotation mode, here the experts performed all annotations without algorithmic support. Finally, there is the expert-algorithm mode where the participants were asked to enhance the generated labels and the artificial flaws. To rule out any memory effects, the participants had a two weeks break between the annotation and the expert-algorithm mode. 

\begin{figure}[hbt!]
\includegraphics[width=1\textwidth]{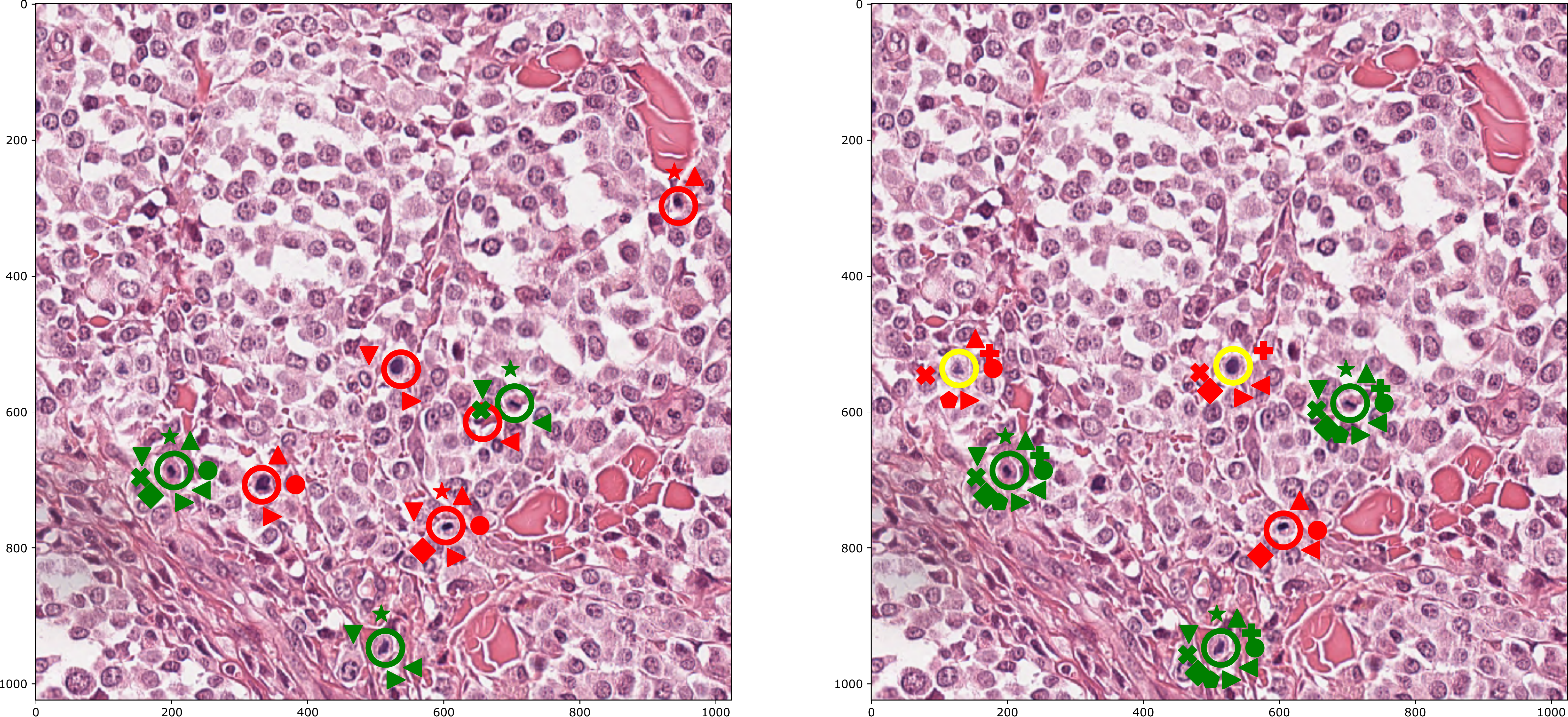}
\caption{Examplary annotation results without (left) and with (right) algorithmic support. Green circles represent ground truth mitotic figures, red false-positives made by experts and yellow fake (artificially introduced) mitotic figures. A symbol at the corresponding annotation represents each expert.}
\label{figSideBySideAnnoAlgo}
\end{figure}

\subsection{VotesLoss}

A key component for training deep neuronal networks is the loss function in combination with the ground truth label weights. In recent years numerous specialised loss functions and label weighting methods have been developed, e.g. focal loss ~\cite{Lin2017ICCV} (to combat class imbalance), label smoothing ~\cite{szegedy2016rethinking} and mixup ~\cite{zhang2017mixup} (both to aid generalisation and robustness). We incorporated the experts' label agreement by assigning a high weight to cells which were annotated by multiple experts using the following two-stage approach:
First, perform a non-maximum suppression for each cell and count how many experts annotated that cell. 
Second, scale the annotations' loss weight function according to Eq. \ref{equationTotalLoss} where $y$ specifies the ground-truth label, $p$ the prediction, $v$ the number of votes, $\mathbf{v}$ the absolute minimal or maximal vote count, and $\alpha$ a scaling factor incorporate additional label smoothing. Finally, the result is multiplied with the cross entropy loss. 

\begin{equation}
\mathrm{VotesLoss}(y,p,v) =-\frac{\alpha}{n} \sum_{i=1}^{n}\left[y_{i} \log \left(p_{i}\right)+\left(1-y_{i}\right) \log \left(1-p_{i}\right)\right] \cdot \frac{v_{i} - \min \mathbf{v}}{ \max \mathbf{v} - \min \mathbf{v}}   
\label{equationTotalLoss}
\end{equation}

To reduce complexity and focus on the loss function, we used a classification and not an object detection task for validation. As a network we used a compact ResNet-18 Architecture~\cite{He:2016ib} pre-trained on ImageNet~\cite{russakovsky2015imagenet}. The network was trained with the Adam optimiser and a maximal learning rate schedule of 0.05. For comparison, we used the binary cross-entropy loss function. As training sets, we used the mitotic figure expert annotations from the annotation mode and the computer-aided mode in two configurations: The single expert set, where a classifier is trained on each expert's annotations, and the number of experts set, where the annotations from two to ten experts are combined. This resulted in a total of 19 training sets for each mode. As validation set, we extracted mitotic figures and non-mitotic figures from the \cite{Bertram_SciData} data set. Each network was trained five times, and the mean of the best results was used for benchmarking.

\section{Results}

\begin{figure}[hbt!]
\includegraphics[width=1\textwidth]{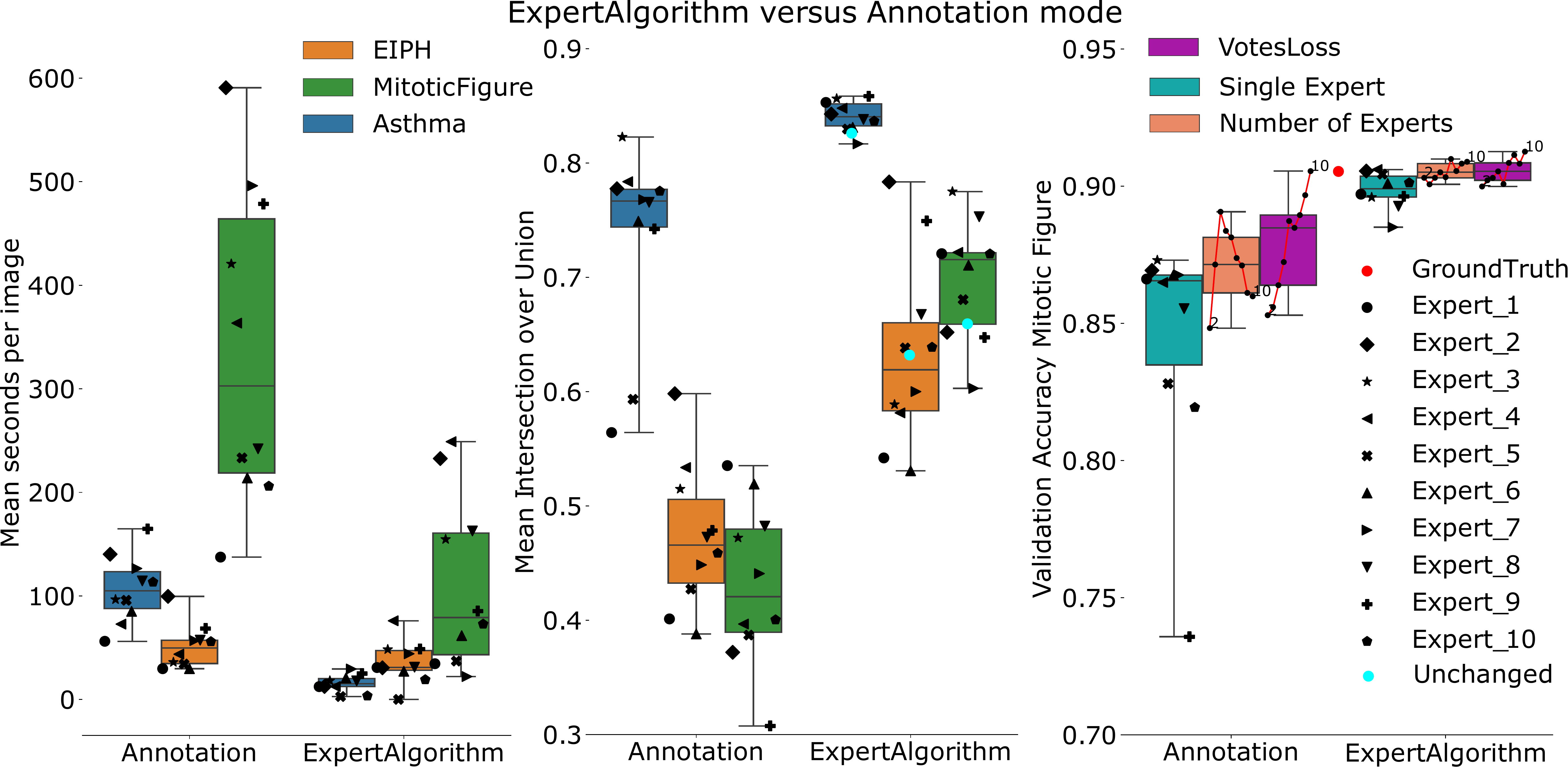}
\caption{Comparison of annotation and expert-algorithm collaboration mode. Left panel shows the mean number of seconds per image, while the middle panel shows the mean intersection over union (mIoU) (threshold of 0.5) for each expert. Additionally, we present the mIoU if all computer-aided annotations were accepted unmodified (Unchanged). On the right, the validation accuracy for the neuronal network trained with data from each expert (Single Expert), with the annotations from two to ten experts combined with a standard loss function (Number of Experts) and with our novel VotesLoss, which incorporates experts annotation agreement, is shown.
}
\label{figResultsAnnoAlgo}
\end{figure}

In total, ten experts made 26,015 annotations on 1,200 images. The annotation accuracy was calculated using objects with mean intersection over union (mIoU) exceeding 0.5. 

\textbf{Deep Learning based Cell Classification:} 
The classifier trained on single expert annotations reached a mean validation accuracy of $\mu$=0.84 (min=0.74, max=0.87, $\sigma$=0.04), for multiple experts ($\mu$=0.87, min=0.85, max=0.89, $\sigma$=0.01) and for multiple experts trained with VotesLoss the accuracy increased to $\mu$=0.88 (min=0.85, max=0.91, $\sigma$=0.02) (see Fig.~\ref{figResultsAnnoAlgo} right). However, in the computer-aided mode, the single experts reached a mean accuracy of $\mu$=0.90 (min=0.89, max=0.90, $\sigma$=0.01), compared to ($\mu$=0.91, min=0.90, max=0.91, $\sigma$=0.01) for multiple experts and ($\mu$=0.91, min=0.90, max=0.91, $\sigma$=0.01) for multiple experts trained with VotesLoss.

\textbf{Asthma:} 
The computer-aided mode led to a significantly increased concordance of $\mu$=0.84 (min=0.82, max=0.86, $\sigma$=0.01) with the ground truth \linebreak[4][F(1,19)=81.61, p$<$0.01], compared to the annotation mode $\mu$=0.73 (min=0.56, max=0.82, $\sigma$=0.08) while also decreasing the annotation time from $\mu$=106 \linebreak[4](min=56, max=164, $\sigma$=30) to $\mu$=15 (min=3, max=29, $\sigma$=11) [F(1,19)=7.17, p=0.01] seconds. The experts found and corrected 78\% of the artificially falsely classified cells, 78\% of the deleted cells and 67\% of the non maximum suppression artefacts in the expert-algorithm mode. In comparison, without pre-annotation of the same cells the experts correctly classified 86\% of the changed classes, 84\% of the deleted cells and 75\% of the non maximum suppression artefacts.   

\textbf{EIPH:} 
The annotation mode ($\mu$=0.47, min=0.39, max=0.60, $\sigma$=0.06) shows significantly decreased concordance with the ground truth \linebreak[4][F(1,19)=42.04, p$<$0.01], compared to the computer-aided mode \linebreak[4]($\mu$=0.59, min=0.53, max=0.67, $\sigma$=0.05) in terms of accuracy and comparable results in terms of annotation time $\mu$=51 (min=29, max=99, $\sigma$=20) to $\mu$=45 (min=27, max=76, $\sigma$=20) seconds. The experts found and corrected 57\% of the artificially falsely classified cells, 10\% of the deleted cells and 54\% of the non maximum suppression artefacts. In comparison without pre-annotation the experts correctly classified 59\% of the changed classes, 28\% of the deleted cells and 60\% of the non maximum suppression artefacts.   

\textbf{Mitotic Figures:} 
The annotation mode ($\mu$=0.43, min=0.31, max=0.54, \linebreak[4]$\sigma$=0.07) shows significantly decreased concordance with the ground truth, compared to the computer-aided mode $\mu$=0.70 (min=0.60, max=0.77, $\sigma$=0.05) in terms of accuracy [F(1,19)=94.71, p$<$0.01]. Annotation time decreases significantly from $\mu$=338 (min=137, max=590, $\sigma$=144) to $\mu$=111 (min=222, max=248, $\sigma$=78) seconds per image for the computer-aided mode [F(1,19)=17.73, p$<$0.01]. The experts corrected 18\% of the artificially removed grade zero cells, 26\% of the grade one cells and 41\% of the grade two cells. They did not remove 71\% of the grade zero fake mitosis, 77\% of grade one and 84\% of grade two. In comparison, without pre-annotation, the experts correctly annotated 26\% grade zero, 43\% grade one and 66\% grade two mitotic figures which were artificially deleted in the computer-aided mode. Furthermore, the experts annotated 14\% of the grade zero fake mitotic figures, 17\% of grade one and 26\% of grade two. According to Elston et al.~\cite{doi:10.1354/vp.46-2-362} seven mitotic figures per high power field is the threshold for grading tumours. In the annotation mode, for six cases experts over-estimated the mitotic figure count, while in 15 cases experts under-estimated the mitotic figure count compared to nine over-estimations and 13 under-estimations in the computer-aided mode.

The analysis code is available online (https://github.com/ChristianMarzahl/ Results-Exact-Study), together with the anonymised participant contributions. The image set is accessible online at https://exact.cs.fau.de/ with the user name: "StudyJan2020" and the password "Alba2020". The study is staying online for further contributions. 

\section{Discussion and outlook}

Our study shows that computer-assisted annotations can lead to a significant improvement regarding the annotation accuracy while also reducing the annotation time, which was a consistent finding in all three experiments. In detail, however, there were differences. \newline
The mitotic figure data set benefited from the computer-aided mode the most with an increase in accuracy of 27\%, which is because mitotic figures are rare and ambiguous to classify. Additionally, we were able to show that our introduced VotesLoss, which incorporated knowledge about how many experts agreed on an annotation, improved the trained classifier accuracy consistently with more annotations from multiple experts in contrast to the standard cross-entropy loss. However, to reach the classification accuracy from one computer-aided expert, up to nine experts with the classical annotation mode would be required. This is not realizable for the creation of large data sets.
Microscopic asthma diagnosis, as a simple and unambiguous task, did not show much of a benefit in terms of accuracy, but the processing speed was increased by a factor of five, which can be attributed to the fact that pathologists were able to check the results at a glance. For EIPH, the picture is slightly different. 
Between the annotation mode and the computer-aided annotated mode no significant time reduction was measurable. We attribute this to the fact that the EIPH grading is more of an estimation that cannot be easily surveyed.\newline 
Overall, we observed that the range of results in the computer-aided mode was reduced, resulting in higher comparability and repeatability of results, which is highly desirable in medicine. Furthermore, we were able to show that it was more likely that the experts overlooked artificially inserted errors (see Fig.~\ref{figSideBySideAnnoAlgo}). At the same time, our results indicate that the systematic errors did not lead to a deterioration of the trained classifier quality. This might be influenced by the artificial labelling errors being introduced symmetrically (i.e., the same number of mitotic figures added as removed), which might have inhibited the creation of an observable bias. Nonetheless, the overlooked artificially inserted errors are a particularly critical observation, as it shows that for all the advantages of speed and accuracy, the quality of the computer-aided annotation is crucial for the result and should be of the highest standard.


\ackname {We thank all contributors for making this work possible. CAB gratefully acknowledges financial support received from the Dres. Jutta \& Georg Bruns-Stiftung f\"ur innovative Veterin\"armedizin.}

%
%
%
%

\bibliographystyle{splncs04}
\bibliography{0000}

\begin{thebibliography}{10}
\providecommand{\url}[1]{\texttt{#1}}
\providecommand{\urlprefix}{URL }
\providecommand{\doi}[1]{https://doi.org/#1}

\bibitem{aubreville2019field}
Aubreville, M., Bertram, C.A., Marzahl, C., Gurtner, C., Dettwiler, M.,
  Schmidt, A., Bartenschlager, F., Merz, S., Fragoso, M., Kershaw, O.,
  Klopfleisch, R., Maier, A.: {Field of Interest Prediction for Computer-Aided
  Mitotic Count}. arXiv.org  \textbf{0}(1902.05414) (2019)

\bibitem{doi:10.1177/0300985819890686}
Bertram, C.A., Aubreville, M., Gurtner, C., Bartel, A., Corner, S.M.,
  Dettwiler, M., Kershaw, O., Noland, E.L., Schmidt, A., Sledge, D.G., Smedley,
  R.C., Thaiwong, T., Kiupel, M., Maier, A., Klopfleisch, R.: Computerized
  calculation of mitotic count distribution in canine cutaneous mast cell tumor
  sections: Mitotic count is area dependent. Vet. Pathol.  \textbf{57}(2),
  214--226 (2020). \doi{10.1177/0300985819890686},
  \url{https://doi.org/10.1177/0300985819890686}, pMID: 31808382

\bibitem{Bertram_SciData}
Bertram, C.A., Aubreville, M., Marzahl, C., Maier, A., Klopfleisch, R.: {A
  large-scale dataset for mitotic figure assessment on whole slide images of
  canine cutaneous mast cell tumor}. Sci. Data  \textbf{6}(274), ~1--9 (2019)

\bibitem{doi:10.1111/vcp.12623}
Bertram, C.A., Dietert, K., Pieper, L., Erickson, N.A., Barton, A.K.,
  Klopfleisch, R.: Effects of on-slide fixation on the cell quality of
  cytocentrifuged equine bronchioalveolar lavage fluid. VET CLIN PATH
  \textbf{47}(3),  513--519 (2018). \doi{10.1111/vcp.12623},
  \url{https://onlinelibrary.wiley.com/doi/abs/10.1111/vcp.12623}

\bibitem{doi:10.1354/vp.46-2-362}
Elston, L.B., Sueiro, F.A., Cavalcanti, J.N., Metze, K.: Letter to the editor:
  The importance of the mitotic index as a prognostic factor for survival of
  canine cutaneous mast cell tumors: A validation study. Vet. Pathol.
  \textbf{46}(2),  362--364 (2009). \doi{10.1354/vp.46-2-362},
  \url{https://doi.org/10.1354/vp.46-2-362}, pMID: 19261652

\bibitem{0000-02}
Ganz, M., Kondermann, D., Andrulis, J., {et al}: Crowdsourcing for error
  detection in cortical surface delineations. Int J Comput Assist Radiol Surg
  \textbf{12}(1),  161--166 (2017)

\bibitem{0000-07}
Golde, D.W., Drew, W.L., Klein, H.Z., {et al}: Occult pulmonary haemorrhage in
  leukaemia. Br Med J  \textbf{2}(5964),  166--168 (1975)

\bibitem{He:2016ib}
He, K., Zhang, X., Ren, S., Sun, J.: {Deep Residual Learning for Image
  Recognition}. In: CVPR. pp. 770--778. IEEE (2016)

\bibitem{Kiupel:2011du}
Kiupel, M., Webster, J.D., Bailey, K.L., Best, S., DeLay, J., Detrisac, C.J.,
  Fitzgerald, S.D., Gamble, D., Ginn, P.E., Goldschmidt, M.H., Hendrick, M.J.,
  Howerth, E.W., Janovitz, E.B., Langohr, I., Lenz, S.D., Lipscomb, T.P.,
  Miller, M.A., Misdorp, W., Moroff, S., Mullaney, T.P., Neyens, I., O'Toole,
  D., Ramos-Vara, J., Scase, T.J., Schulman, F.Y., Sledge, D., Smedley, R.C.,
  Smith, K., {W Snyder}, P., Southorn, E., Stedman, N.L., Steficek, B.A.,
  Stromberg, P.C., Valli, V.E., Weisbrode, S.E., Yager, J., Heller, J., Miller,
  R.: {Proposal of a 2-Tier Histologic Grading System for Canine Cutaneous Mast
  Cell Tumors to More Accurately Predict Biological Behavior}. Vet. Pathol.
  \textbf{48}(1),  147--155 (jan 2011). \doi{10.1177/0300985810386469}

\bibitem{Lin2017ICCV}
Lin, T.Y., Goyal, P., Girshick, R., He, K., Doll{\'a}r, P.: Focal loss for
  dense object detection. In: ICCV. pp. 2980--2988 (Oct 2017)

\bibitem{0000-01}
Maier-Hein, L., Ross, T., Gr{\"o}hl, J., {et al}: Crowd-algorithm collaboration
  for large-scale endoscopic image annotation with confidence. In: Med Image
  Comput Comput Assist Interv. pp. 616--623. Springer (2016)

\bibitem{0000-10}
Marzahl, C., Aubreville, M., Bertram, C.A., Gerlach, S., Maier, J., Voigt, J.,
  Hill, J., Klopfleisch, R., Maier, A.: Fooling the crowd with deep
  learning-based methods. arXiv preprint arXiv:1912.00142  (2019)

\bibitem{Meyer:2005cl}
Meyer, J.S., Alvarez, C., Milikowski, C., Olson, N., Russo, I., Russo, J.,
  Glass, A., Zehnbauer, B.A., Lister, K., Parwaresch, R.: {Breast carcinoma
  malignancy grading by Bloom-Richardson system vs proliferation index:
  Reproducibility of grade and advantages of proliferation index}. Modern
  Pathology  \textbf{18}(8),  1067--1078 (aug 2005).
  \doi{10.1038/modpathol.3800388}

\bibitem{0000-00}
{\O}rting, S., Doyle, A., van Hilten, M.H.A., {et al}: A survey of
  crowdsourcing in medical image analysis. arXiv preprint arXiv:1902.09159
  (2019)

\bibitem{russakovsky2015imagenet}
Russakovsky, O., Deng, J., Su, H., Krause, J., Satheesh, S., Ma, S., Huang, Z.,
  Karpathy, A., Khosla, A., Bernstein, M., et~al.: Imagenet large scale visual
  recognition challenge. Int J Comput Vis  \textbf{115}(3),  211--252 (2015)

\bibitem{szegedy2016rethinking}
Szegedy, C., Vanhoucke, V., Ioffe, S., Shlens, J., Wojna, Z.: Rethinking the
  inception architecture for computer vision. In: CVPR. pp. 2818--2826 (2016)

\bibitem{zhang2017mixup}
Zhang, H., Ciss{\'{e}}, M., Dauphin, Y.N., Lopez{-}Paz, D.: mixup: Beyond
  empirical risk minimization. In: ICLR. OpenReview.net (2018),
  \url{https://openreview.net/forum?id=r1Ddp1-Rb}

\end{thebibliography}

\end{document}